\documentclass[12pt]{article}
\usepackage{epsfig}

\newcommand{\psb}{\bar{\psi}}

\newcommand{\eqn}[1]{Eq.(\ref{#1})}
\newcommand{\gev}{\mbox{ GeV}}

\newcommand{\hl}{\hline}
\newcommand{\ben}{\begin{enumerate}}
\newcommand{\een}{\end{enumerate}}
\newcommand{\bit}{\begin{itemize}}
\newcommand{\eit}{\end{itemize}}
\newcommand{\bc}{\begin{center}}
\newcommand{\ec}{\end{center}}
\newcommand{\bq}{\begin{equation}}
\newcommand{\eq}{\end{equation}}
\newcommand{\bqa}{\begin{eqnarray}}
\newcommand{\eqa}{\end{eqnarray}}
\newcommand{\lb}{\linebreak}
\newcommand{\nn}{\nonumber}
\newcommand{\mtt}[1]{\mbox{\tt{#1}}}
\newcommand{\plaat}[3]{\raisebox{#3pt}{\epsfig{figure=#1.eps,
width=#2cm}}}

\newcommand{\slaa}[2]{#1\hspace*{#2pt}/}

\def\demo{$\Delta\eta\mu \acute{o} \kappa \varrho \iota \tau o \varsigma$}

\begin{document}

\pagestyle{empty}

\begin{flushright}
DEMO-HEP-2000/01 \\
\end{flushright}

\vspace*{2cm} 

\bc\begin{LARGE} {\bf HELAC: a package to compute
electroweak helicity amplitudes }\\ \end{LARGE}
\vspace*{2cm}

{\large {\bf Aggeliki Kanaki}  and {\bf Costas G.~Papadopoulos} } \\[12pt]
Institute of Nuclear Physics, NCSR \demo, 15310 Athens, Greece
\\

\vspace*{1cm}

{\bf ABSTRACT}\\[12pt]  \ec

\begin{quote}

\mtt{HELAC} is a {\tt FORTRAN} based package that is able to compute
efficiently helicity amplitudes for arbitrary scattering processes
within the standard electroweak theory. The algorithm
exploits the virtues of the Dyson-Schwinger equations as compared
to the traditional Feynman graph approach. All electroweak
vertices are included in both the unitary and Feynman gauges, and
computations including all mass effects are available. A version performing
multi-precision computations with arbitrary - user defined -
accuracy is also included, allowing access to any phase space
point for arbitrary high energies.

\end{quote}

\vspace*{\fill}

February 2000

\vspace*{\fill}





\newpage
\pagestyle{plain}


\bc {\bf PROGRAM SUMMARY}\\[18pt]\ec
{\it Title of the program:} \\
{\tt HELAC}. \\[8pt]
{\it Catalogue number:}  \\[8pt]
{\it Program obtainable from:}\\
 Dr. Costas G. Papadopoulos, 
Institute of Nuclear Physics,
NRCPS `Democritos', 15310 Athens, Greece, \mtt{e-mail} 
\mtt{Costas.Papadopoulos@cern.ch}.\\
Also at {\tt ftp://alice.nuclear.demokritos.gr/dist/pub/papadopo/helac}.\\[8pt]
{\it Licensing provisions:} none\\[8pt]
{\it Computer for which the program is designed and others on 
which it has been tested:}\\
ALPHA, HP, IBM workstations.\\[8pt]
{\it Operating system under which the program has been tested:}\\ 
Unix.\\[8pt]
{\it Programming language:} \\
FORTRAN 77 and FORTRAN 90\\[8pt]
\\[8pt]
{\it Nature of physical problem:}\\ 
A substantial part of particle phenomenology nowadays is based
upon our ability to study efficiently processes involving a
relatively large number of particles. This requires efficient
algorithms for matrix element calculation and phase space
generation and integration. As far as the matrix element
calculation is concerned, the traditional approach utilizes
standard Feynman graph representation of the scattering amplitude,
resulting to a computational cost that grows asymptotically as
$n!$, where $n$ is the number of particles involved in the
process. 
\\[8pt]
{\it Method of solution:} \\
As an alternative recursive algorithms based on
Schwinger-Dyson equations lead asymptotically to a much lower
growth of the computational cost, namely
$a^n$, where $a \sim 3$. It is the aim of this paper to present
such an algorithm as well as the corresponding {\tt FORTRAN} code
which allows the computation of any tree-order electroweak
amplitude.
\newpage




\section{Introduction}

A substantial part of current particle phenomenology is based
upon our ability to study efficiently processes involving a
relatively large number of particles. This requires efficient
algorithms for matrix element calculation and phase space
generation and integration. As far as the matrix element
calculation is concerned, the traditional approach utilizes
standard Feynman graph representation of the scattering amplitude,
resulting to a computational cost that grows asymptotically as
$n!$, where $n$ is the number of particles involved in the
process. As an alternative recursive algorithms based on
Schwinger-Dyson equations lead asymptotically to a much lower
growth of the computational cost, namely
$a^n$, where $a \sim 3$. It is the aim of this paper to present
such an algorithm as well as the corresponding {\tt FORTRAN} code
which allows the computation of any tree-order electroweak
amplitude.


%


\section{The algorithm}

\par

As advertised, the algorithm is based upon the well-known Dyson-Schwinger
equations. These equations are equivalent to the
field equations derived from the classical lagrangian and
express recursively the $n$-point Green's functions
in terms of the $1-,2-,\ldots,(n-1)$-point functions.

In order to illustrate this point let us consider a simple example
of a QED-like theory possessing minimal interactions of a spinor
field to a gauge boson. Let $p_1,p_2,\ldots,p_n$ represent the
external momenta involved in the scattering process under
consideration, taken to be incoming. For a vector field we define
\bq b_\mu(P)\;\;=\;\; \plaat{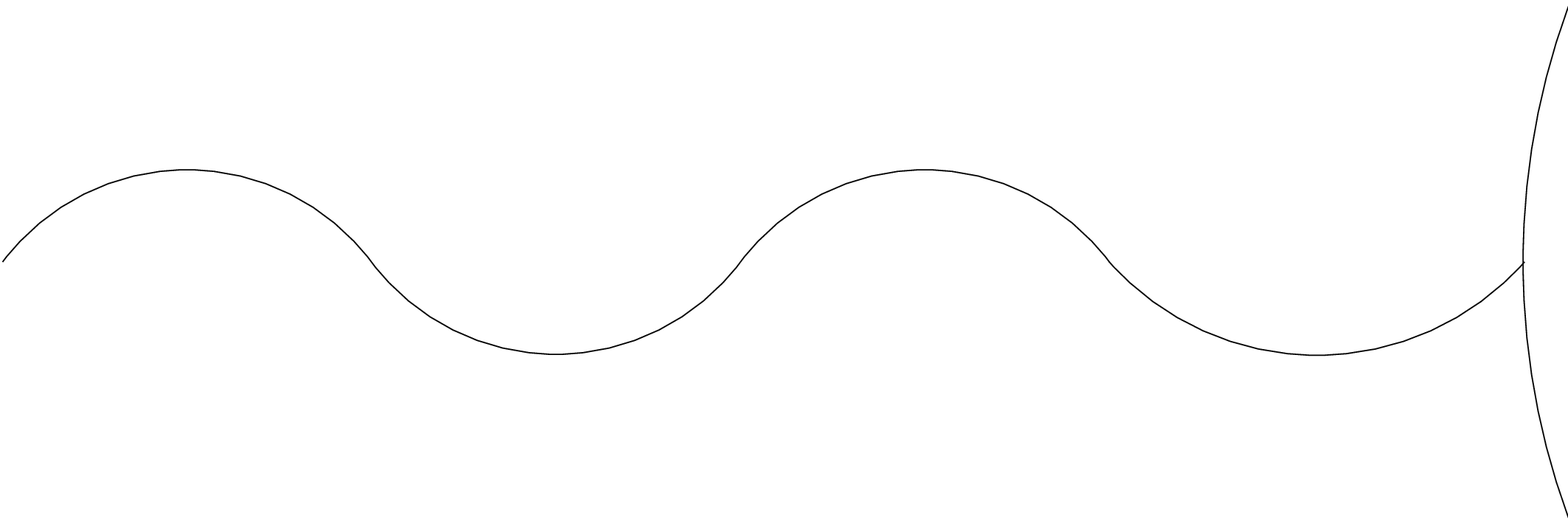}{2}{-12} \eq a four vector,
which describes any sub-amplitude from which a boson $V$ with
momentum $P$ can be constructed. The momentum $P$ is given as a
sum of  momenta of external particles, namely $P=\sum_{i\in I}
p_i$ with $I\subset \{1,\ldots,n\}$. Accordingly, we define by \bq
\psi(P)\;\;=\;\; \plaat{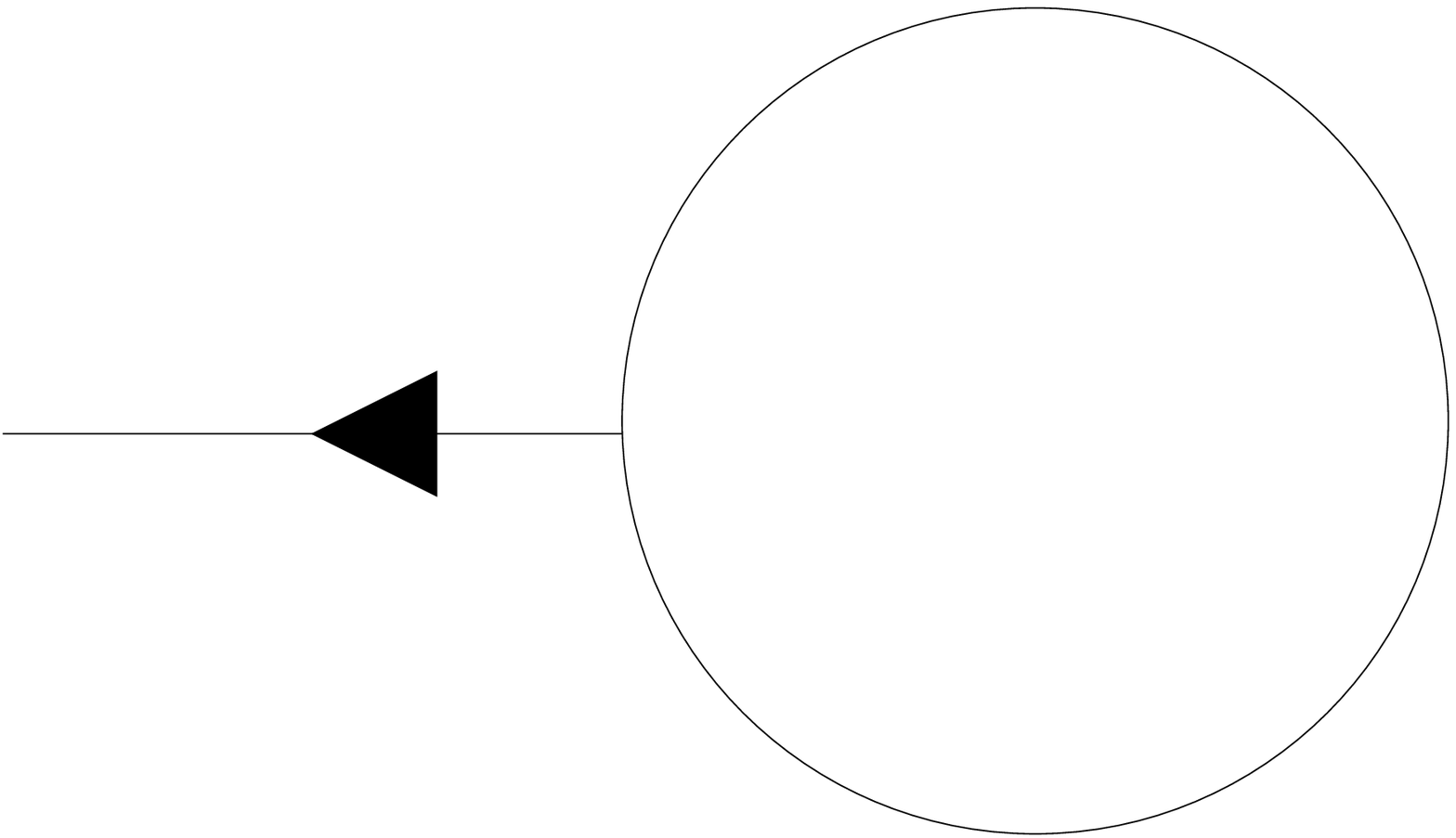}{2}{-12} \eq a four-dimensional
spinor, which describes any sub-amplitude from which a fermion
with momentum $P$ can be constructed and by \bq \psb(P)\;\;=\;\;
\plaat{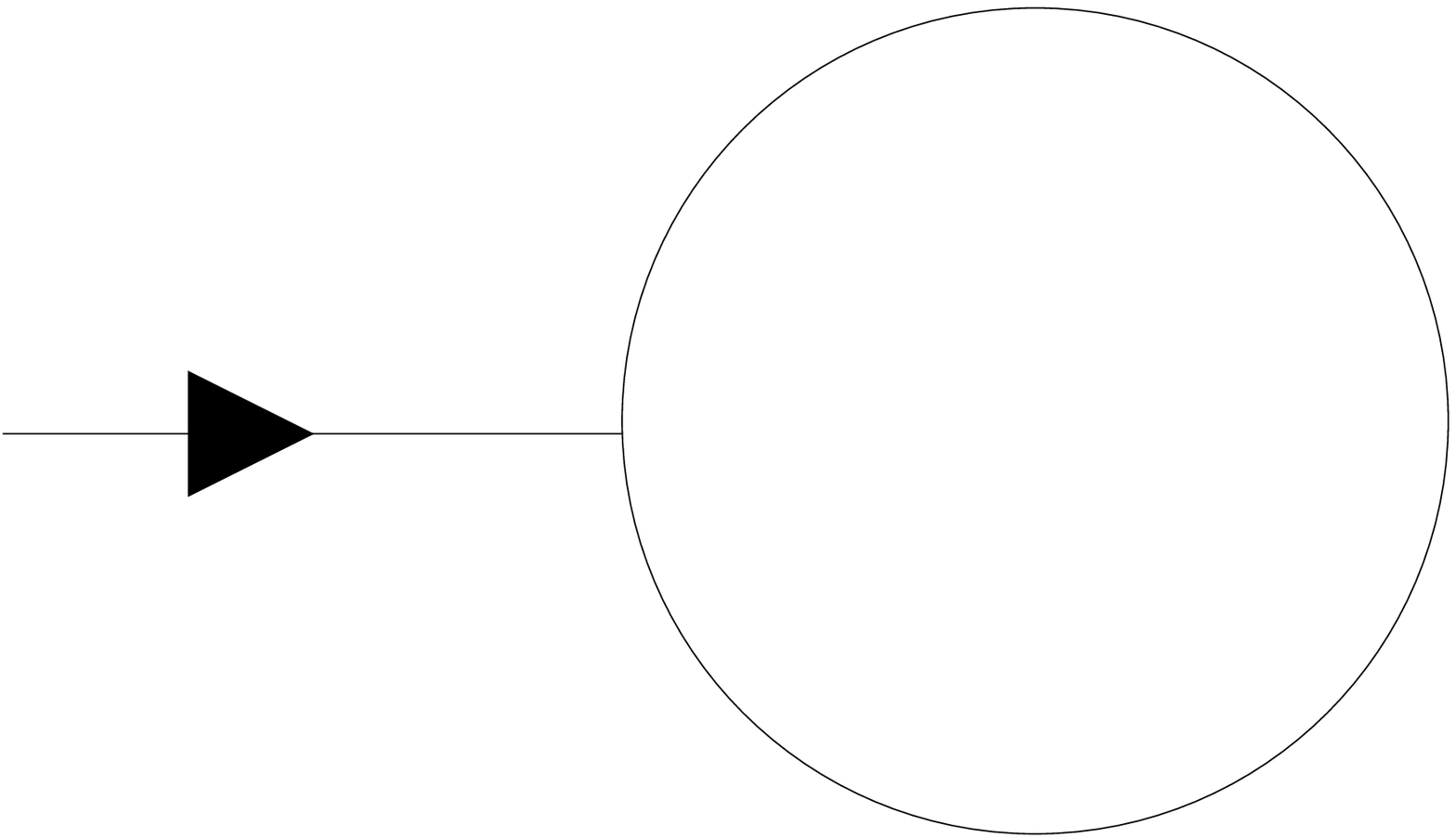}{2}{-12} \eq a four-dimensional antispinor, which
describes any sub-amplitude from which an antifermion with
momentum $P$ can be constructed.

The Dyson-Schwinger equations take the following simple form. For
a boson with momentum $P^\mu$, 
\bqa
\plaat{bos1}{2}{-12}\;\;&=&\;\plaat{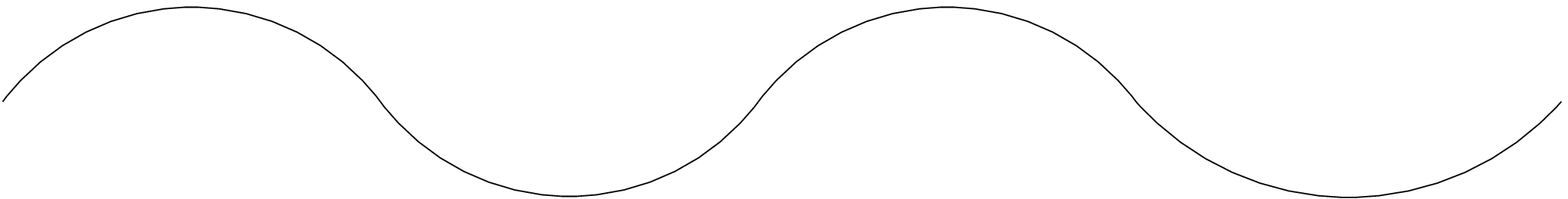}{1.5}{-2}\;
+\;\plaat{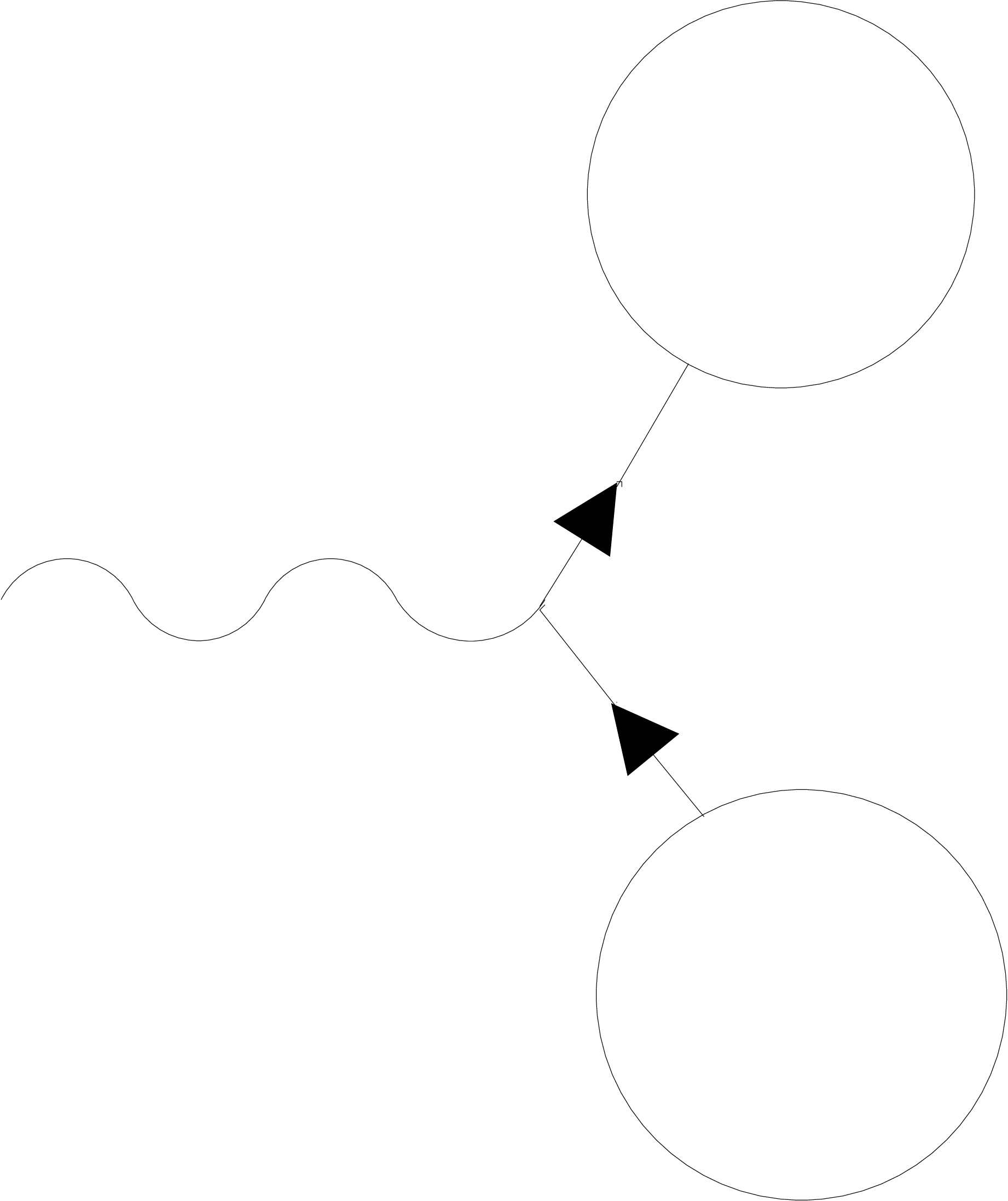}{2}{-32} 
\nn
\\ 
b^\mu(P)&=&\sum_{i=1}^n
\delta_{P=p_i} b^\mu(p_i) + \sum_{P=P_1+P_2}
\;(ig)\;\Pi^\mu_\nu\; \psb(P_2)\gamma^\nu\psi(P_1)
\epsilon(P_1,P_2) \label{vff} 
\eqa 
where the propagator $\Pi$ is
given by 
\[ \Pi^\mu_\nu=\frac{i}{P^2-m^2}
\biggl(-g^\mu_\nu+\frac{P^\mu P_\nu}{P^2-\xi m^2}(1-\xi)\;.
\biggr) 
\] 
and $\epsilon(P_1,P_2)$ is a sign factor, that takes
the value $\pm 1$ and whose definition will be described below.
For a fermion with momentum $P$ \bqa
\plaat{afer1}{2}{-12}\;\;&=&\plaat{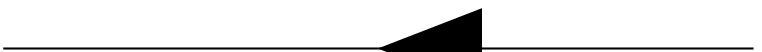}{1.5}{3}\; +
\;\plaat{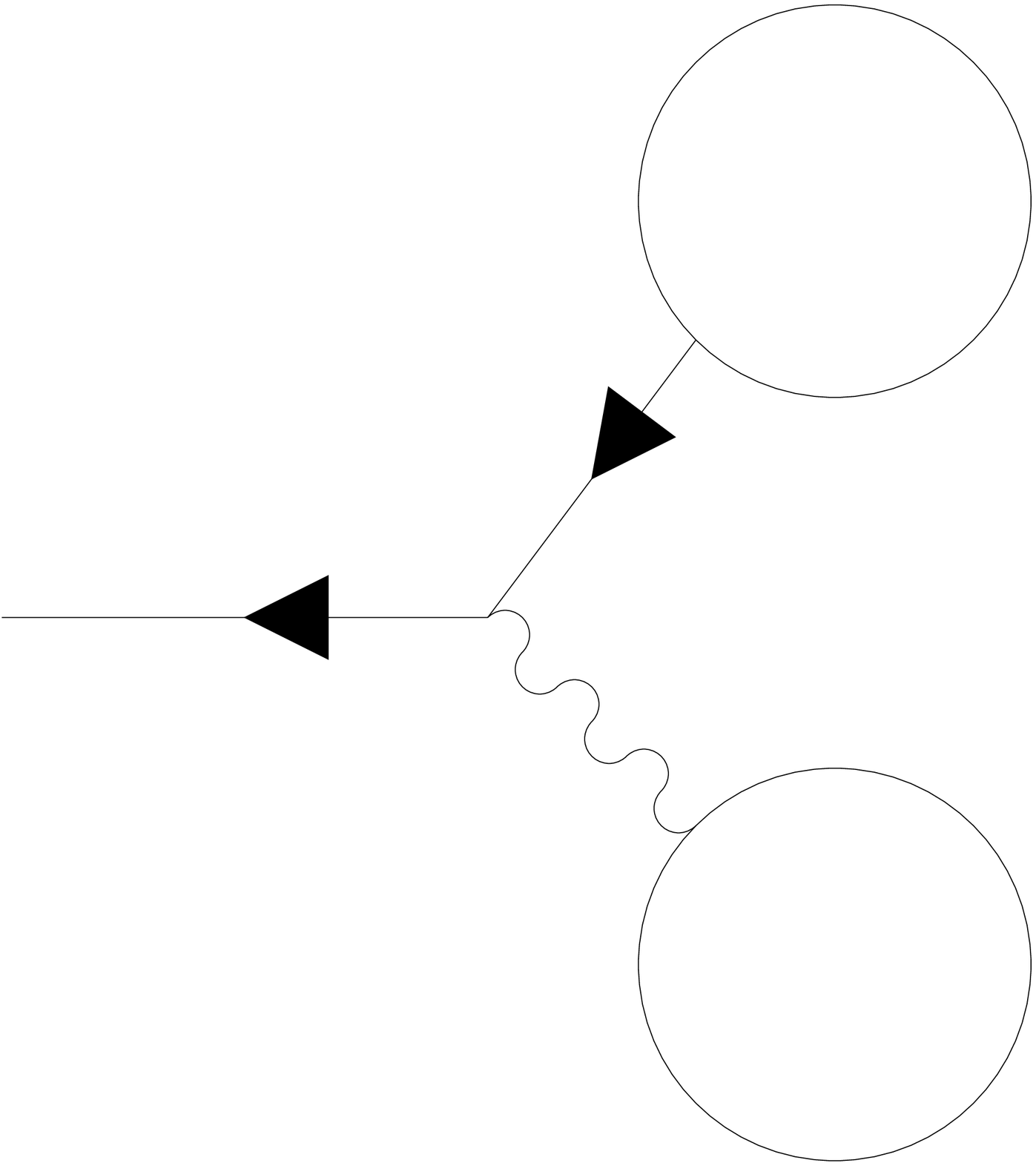}{2}{-27} 
\nn
\\ 
\psi(P)&=&\sum_{i=1}^n \delta_{P=p_i}
\psi(p_i) + \sum_{P=P_1+P_2} (ig) {\cal
P}\;\slaa{b}{-5.5}(P_2)\psi(P_1)\; \epsilon(P_1,P_2)
\label{ffv} \eqa 
where ${\cal P}$ is given by 
\[ {\cal
P}=\frac{i}{\slaa{P}{-8}-m} 
\]
 and for an antifermion with
momentum $P$, \bqa
\plaat{fer1}{2}{-12}\;\;&=&\;\plaat{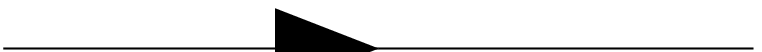}{1.5}{3}\;+
\;\plaat{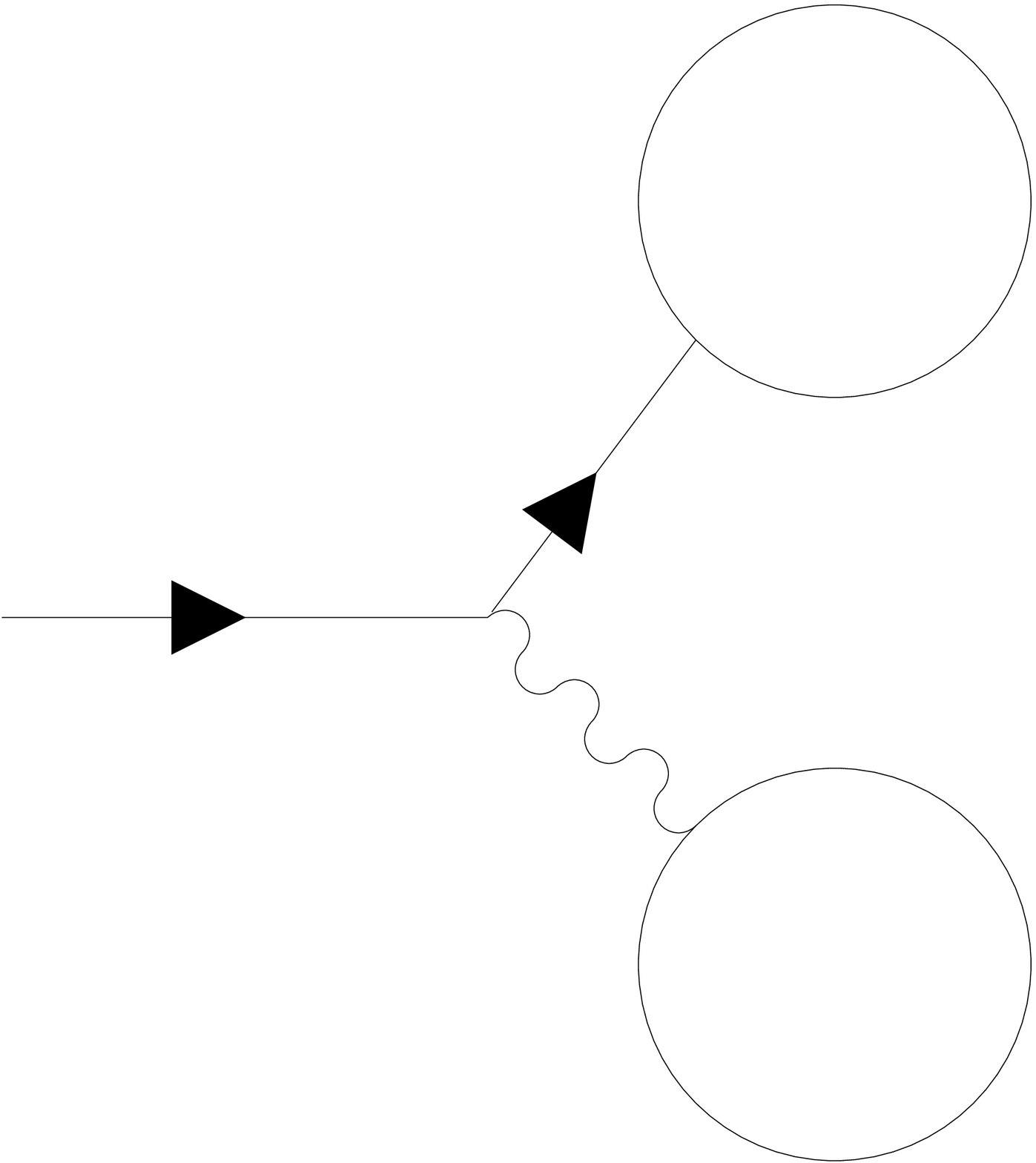}{2}{-27} 
\nn
\\ 
\psb(P)&=&\sum_{i=1}^n \delta_{P=p_i}
\psb(p_i) +\sum_{P=P_1+P_2} (ig) \psb(P_1)\slaa{b}{-5.5}(P_2) \bar{{\cal
P}} \;\epsilon(P_1,P_2) 
\label{affv} \eqa 
where 
\[ \bar{{\cal
P}}=\frac{i}{-\slaa{P}{-8}
-m}\;. 
\]

The scattering amplitude can be calculated by any of the following
relations,
\bq
{\cal A}(p_1,\ldots,p_n) = \left\{
\begin{array}{ll}
b_0^\mu(P_i) b_\mu(p_i) & \mbox{where $i$ corresponds to a photon}
\\ \bar{\psi}_0(P_i){\psi}(p_i) & \mbox{where $i$  corresponds to
an incoming fermion line}  \\ \psb(p_i)\psi_0(P_i) & \mbox{where
$i$  corresponds to an outgoing fermion line}  \\
\end{array}
\right.
\eq
where
\[ P_i=\sum_{j\not= i}p_j,\]
so that $P_i+p_i=0$. The functions with  subscript $0$ are given by the 
previous expressions,
except for the propagator term. This is because the outgoing momentum $P_i$
being on shell the propagator factor is removed by the amputation procedure.
The initial conditions are given by
\bqa
b^\mu(p_i)&=&\epsilon^\mu_\lambda(p_i) ,    \lambda=\pm1,0
\nn\\
\psi(p_i)&=&\left\{
\begin{array}{ll}
u_\lambda(p_i)
&\mbox{if $p_i^0\geq0$} \\
v_\lambda(-p_i)
&\mbox{if $p_i^0\leq0$} \\
\end{array}
\right.
\nn\\
\psb(p_i)&=&\left\{
\begin{array}{ll}
\bar{u}_\lambda(p_i)
&\mbox{if $p_i^0\geq0$} \\
\bar{v}_\lambda(-p_i)
&\mbox{if $p_i^0\leq0$} \\
\end{array}
\right.
\label{ampl}\eqa
where the explicit form of $\epsilon^\mu_\lambda,u_\lambda,
v_\lambda,\bar{u}_\lambda,\bar{v}_\lambda$
are given in the Appendix.

In order to actually solve the recursive equations it is convenient to use
a binary representation of the momenta involved~\cite{camo}.
For a process involving $n$ external particles with momenta 
$p_i^\mu,i=1\ldots,n$
we assign to the momentum $P^\mu$ defined as
\[
P^\mu=\sum_{i\in I}p_i^\mu
\]
where $I\subset \{ 1,\ldots,n\}$, a binary vector $\vec{m}=(m_1,\ldots,m_n)$,
where its components take the values $0$ or $1$ in such a way that
\[
P^\mu=\sum_{i=1}^n m_i\;p_i^\mu\;.
\]
Moreover this binary vector can be uniquely represented by the
integer
\[
m=\sum_{i=1}^n 2^{i-1}m_i
\]
with  $0\le m \le 2^{n-1}$. Especially, the external momenta
$p_i$, $i=1\ldots,n$ are represented by $m=2^{i-1},i=1,\ldots,n$.
All momenta $P$ can now be replaced by the corresponding integers
\[ b_\mu(P)\to b_\mu(m)\;.\]
A very convenient ordering of integers in binary representation
relies on the notion of level $l$, defined simply as
\[
l=\sum_{i=1}^n m_i\;.
\]
As it is easily seen all external momenta are of level $1$, whereas
the total amplitude corresponds to the unique level $n$ integer
$2^{n-1}$. This ordering dictates the natural path of the
computation; starting with level-$1$ sub-amplitudes, we compute
the level-$2$ ones using the Dyson-Schwinger equations and so on
up to the final expression \eqn{ampl}. As far as the sign factor
is concerned
\[ \epsilon(P_1,P_2) \to \epsilon(m_1,m_2) \]
we define 
\bq \epsilon(m_1,m_2)=(-1)^{\chi(m_1,m_2)} \eq with \bq
\chi(m_1,m_2)= \sum_{i=n}^2 \hat{m}_{1i}\left( \sum_{j=1}^{i-1}
\hat{m}_{2j}\right) 
\label{signfactor}\eq 
where hated components are set to $0$ if
the corresponding external particle is a boson. It is not
difficult to see that this sign factor takes properly into account
the anti-symmetry of the amplitude with respect to fermionic
particles.

As an illuminating example we present here the computation of the
amplitude for the process $e^+e^-\rightarrow\mu^+\mu^-\gamma$. The
calculation starts with the computation of the level-$2$
sub-amplitudes, which are all sub-amplitudes produced from any two
of the external particles. These are \bqa
b_\mu(12)&=&(ig)\Pi_{12\mu}^\nu\psb(4)\gamma_\nu\psi(8) \nn\\
\psb(18)&=&(ig)\psb(2)\slaa{b}{-5.5}(16)\bar{{\cal P}}_{18} \nn\\
\psb(20)&=&(ig)\psb(4)\slaa{b}{-5.5}(16)\bar{{\cal P}}_{20}\nn\\
\psi(24)&=&(ig){\cal P}_{24}\slaa{b}{-5.5}(16)\psi(8) \nn\eqa Then
we compute  the level-$3$ sub-amplitudes: \bqa
\psb(14)&=&(ig)\psb(2)\slaa{b}{-5.5}(12)\bar{{\cal P}}_{14}\nn\\
b_\mu(28)&=&(ig)\Pi_{28\mu}^\nu \left(\psb(20)\gamma_\nu\psi(8)
 +\psb(4)\gamma_\nu\psi(24)\right)
\nn\eqa and then the level-$4$ sub-amplitude, which is the final
level in this case, 
\bq \psb_0(30)=(ig)\left( \psb(2) \slaa{b}{-5.5}(28)
         + \psb(14)\slaa{b}{-5.5}(16)
         + \psb(18)\slaa{b}{-5.5}(12)\right)
\eq
Then the amplitude is simply given by
\[
{\cal A}=\psb_0(30)\psi(1)
\]
Note that we have chosen the particle number $1$ as our ending
point, so we have computed all sub-amplitudes where the momentum
$p_1$ does not appear: this excludes all odd integers between $1$
and $2^{n-2}$.

There are two special issues in the computation which go beyond the
recursive equations presented above. The first is the generation
of all helicity configurations and the second the treatment
of the color summation in the case external particles with color
are involved.

As far as the helicity configurations are concerned this is done in a rather
straightforward way, resulting to an automatic evaluation
of all relevant combinations. For any given external particle
knowledge of its flavor allows the program to compute all relevant
helicity configurations. The total number is given by
$ 2^{l_2} 3^{l_3}$ where $l_2$ is the number of (anti-)fermions and massless
gauge bosons and $l_3$ the number of massive gauge bosons involved
in the scattering process.

The color configurations are taken into account as follows. Since
only electroweak processes are considered in this version, the
only colored particles that can appear in the amplitude are quarks
and antiquarks and let us have $n$ pairs of them. Each color
amplitude is proportional to the following color structure

\[
{\cal C}_i=\delta_{1,\sigma_i(1)}\delta_{2,\sigma_i(2)}
\ldots\delta_{n,\sigma_i(n)}
\]

where  $\sigma_i$ represents the $i$-th permutation of the set
${1,2,\ldots,n}$. The code computes all non-vanishing color
amplitudes as well as the corresponding color matrix
\[
{\cal M}_{ij}=\sum C_i C_j
\]
and finally performs the color summation.




\section{The code {\tt HELAC}}

The computation is split in two major phases. During the first
phase, which we call {\it initialization phase}, the program selects all
the relevant sub-amplitudes for the required process. In the
simplest version the program accepts as input the following
variables: \bit
\item {\tt n} the number of particles involved in the scattering
\item {\tt ifl(1:n)} the array of flavors of the particles
\item {\tt iflag}. If set to \mtt{0} sum over all helicity configurations is
understood. If set to \mtt{1} you must supply also the specific helicity 
configuration to be computed.
\item {\tt iunitary}. If set to $0$ the Feynman gauge is considered whereas if
set to $1$ the unitary gauge is used.
\item {\tt ihiggs} denotes the inclusion (\mtt{1}) or not  (\mtt{0}) of the 
Higgs particle as an intermediate state.
\item {\tt iwidth} denotes the fixed (\mtt{0}) or complex (\mtt{1}) scheme
for the introduction of the width of $W$ and $Z$\footnote{For a recent
discussion see reference~\cite{chap}.}.
\item {\tt io(1:n)} is used to distinguish among initial (\mtt{1})
and final state particles (\mtt{-1}). By default \mtt{io(1:2)=1}
and \mtt{io(3:n)=-1}.
\eit
 The first routine called after the input has been read is
the routine {\tt physics/physics.f} located in the homonymous
file. In this routine all couplings of the standard electroweak
theory are defined~\cite{denn}.
 
Then the routine {\tt helac\_init/master.f} is called. In the
beginning the average-over-helicity ({\tt avhel}), the
average-over-color ({\tt avcol}) and the symmetry({\tt symet})
factors are computed. Then, by calling the routine {\tt
setncc/intpar.f} the number of color configurations {\tt ncc} is
set up depending on the number of quarks and anti-quarks involved.
For each color configuration the program constructs the skeleton
of the amplitude starting at level two and proceeding up to level
$n-1$. All possible vertices are scanned in this phase, as for
instance {\tt vff/pan1.f} which describes the coupling of a
fermion anti-fermion to produce a vector boson, in order to select
all non-zero sub-amplitudes. A special routine,{\tt redo/pan1.f}
is called at the end just to check whether all the selected
sub-amplitudes are indeed contributing to the final amplitude
under consideration.  The program ends this phase when all color
configurations have been considered and the color matrix {\tt
rmatrix(1:ncc,1:ncc)} has been computed. 

During the second phase, which we call the {\it computation phase}, 
the code computes
numerically the amplitude for each phase space point introduced.
The main calling routine is  \lb {\tt helac\_master/master.f}. The first step
is the automatic setup of the helicity configurations via the routine
{\tt sethel/intpar.f}. The next step is the computation of the 
external particle
wave functions by the routine {\tt iniqq/pan1.f}. All relevant routines
for this computation can be found in {\tt wavef.f}. The program proceeds by
computing the amplitude for every color configuration via {\tt nextq/pan1.f}.
The vertex and propagator functions needed are included in {\tt pan2.f}.
Moreover the sign factor as described in \eqn{signfactor} is also computed.
The routine {\tt helac\_master/master.f} returns the squared matrix element
({\tt smel2})summed and averaged over helicity and color.
Finally, a sample {\tt main/main.f} program to run {\tt HELAC} is provided.
In this part of the code a call to a phase-space generator
is also included ({\tt rambo/rambo.f}), in order to generate
appropriate phase space points for the computation.

All floating point computations are performed in {\tt real*8} or
double  precision. Nevertheless the program is written in such a
way that one can choose a higher accuracy if needed. There are two
alternatives. The first one is to use {\tt real*16} or quadruple
precision. For this task several modules to perform quadruple
precision computation with complex numbers are supplied in {\tt
qprec.f}. This is necessary because in most of the existing
platforms no {\tt FORTRAN} quadruple precision complex numbers are
available. The most efficient way to use this precision with
complex numbers is to exploit the virtues of {\tt FORTRAN 90} and
define the appropriate derived types and modules. The second
alternative make use of a multi-precision library~\cite{mpcl}, 
written also in
{\tt FORTRAN 90}. The precision is now user-defined and is set up by
calling the routine {\tt zmset} in the very beginning of the
program ({\tt main.f}). All these are automatically driven by the
appropriate {\tt make} files included in the package.




\section{Results}

In this section we will try to explain how to read the output of the
program and to highlight several aspects of it. First of all
let us follow a sample computation of the process
\[
e^- e^+ \to  e^- e^+ \gamma\;.
\]
As an input we have to provide 
\ben
\item
The number of particles involved in the process (\mtt{5}).
\item For each particle its flavor (\mtt{2 -2 2 -2 31}).
In the following table we provide the correspondence used in \mtt{HELAC}:\\
\bc\begin{tabular}{|c|c|}
\hl
$\nu_e,e^-,u,d,\nu_\mu,\mu^-,c,s,\ldots$ & $1,\ldots,12$ 
\\
$\bar{\nu_e},e^+,\bar{u},\bar{d},\bar{\nu}_\mu,\mu^+,
\bar{c},\bar{s},\ldots$ & $-1,\ldots,-12$ \\
$\gamma,Z,W^+,W^-$ & $31,\ldots,34$ \\
$H,\chi,\phi^+,\phi^-$ & $41,\ldots,44$ \\
\hl 
\end{tabular}\ec
\item The parameters \mtt{iflag}, \mtt{iunitary}, \mtt{ihiggs},
\mtt{iwidth} (\mtt{0 1 0 0}).
\een
The expected output is as follows
\\
\rotatebox{-90}{\plaat{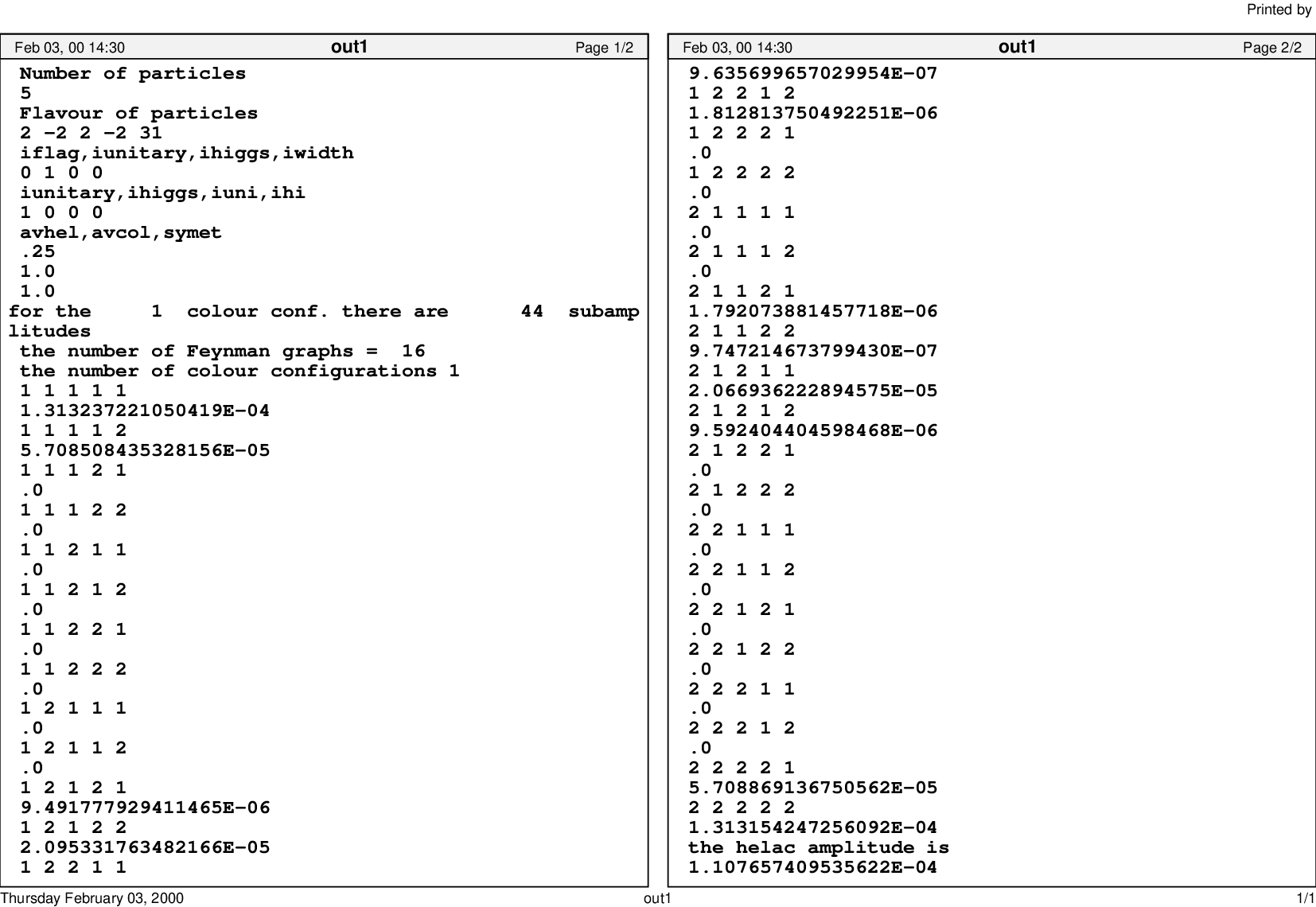}{11}{0}}
\\[12pt]
where the first lines show the required input data; then the
average-helicity, average-color and symmetry factors are shown;
the number of contributing sub-amplitudes for each color
configuration is given ({\tt 44}) as well as the total number of
Feynman graphs ({\tt 16}). This marks the end of the
initialization phase, and afterwards the computation of the
squared matrix element for a given phase space point is performed.
Results for each helicity configuration are printed and at the end
the {\tt HELAC} amplitude squared is given ({\tt
1.107657409535622E-04})

As far as the efficiency of the code is concerned extensive 
comparisons have been made to existing calculations. We have chosen
among them two popular tools, namely  \mtt{MADGRAPH}~\cite{madg} 
and \mtt{EXCALIBUR}~\cite{exca}.
Comparisons with \mtt{EXCALIBUR} have been restricted to four-fermion
final states . Running under the same conditions the speed ratio
was varying between $1$ and $2$. The same results have been obtained
in comparisons with \mtt{MADGRAPH} taken into account that
it can produce results for only up to $7$ particles involved
in the scattering process. As expected \mtt{HELAC} show an exponential
(instead of factorial) \mtt{CPU}-time growth. There is no a priori
limitation for the number of particles that \mtt{HELAC} can treat,
the only restrictions being that of memory allocation. 
 
In order to have a taste of a multi-precision computation we have
computed the squared amplitude for the process
\[
e^-(p_1) e^+(p_2) \to e^-(p_3) e^+(p_4) e^-(p_5) e^+(p_6)
\]
at two phase space points. Phase space point (A) is just a
randomly generated one by the phase-space generator {\tt RAMBO}.
Phase space point (B) on the other hand is a very special one,
where 
\bqa &p_1^0/\gev=100
,\;\;\vec{p_1}+\vec{p_2}=0,\;\;p_3^0/p_1^0=0.9,
\;\;\theta_3=0,& \nn\\ &(p_5+p_6)^2/(p_4+p_5+p_6)^2=0.1,\;\;
\theta_4=0, \;\;\phi_4=0,\;\; \theta_5=0,\;\; \phi_5=0& \nn \eqa
and $m_e=0.511\times 10^{-3}\gev$.
In this case all particles are co-linear to the beam
and therefore important enhancements are expected. In the following table
results are provided for these points, by using the {\tt real*8}
(DP), {\tt real*16} (QP) and 34-digit multi-precision version of
the code.
\\[12pt]
\begin{tabular}{ll}
\parbox{5cm}{\bc (A)\ec} &\parbox{5cm}{\bc (B)\ec}  \\
{\footnotesize 1.539728523150595E-008}
&
{\footnotesize 1.256276706229023E+023}
\\
{\footnotesize 1.53972852315058854156763002825013D-08}
&
{\footnotesize 3.07162601093710915134136924973089D+22}
\\
{\footnotesize 1.53972852315058854156763002825011853M-8}
&
{\footnotesize 3.07162601093710915127950109241770808M+22}
\end{tabular}\\[12pt]
As one can easily see the numerical stability of the DP
computation is spoiled in the co-linear region. This is due to
huge gauge cancellations occurring in this region; a way out of
this problem will be discussed elsewhere.
 
\vspace*{1cm}

We end this presentation by summarizing the main achievements:

\bit
\item An algorithm based on Dyson-Schwinger equations has been presented
that enables the computation of electroweak amplitudes with high
efficiency.
\item Based on this algorithm a {\tt FORTRAN} package {\tt HELAC} has been
developed, which includes full massive computation in both the
unitary and the Feynman gauge of the standard electroweak theory.
\item A  quadruple as well as a  multi-precision version of {\tt HELAC} have
 been incorporated allowing numerically stable results for any phase space
point and for arbitrary high energies.
\eit


%






\noindent{\Large\bf Appendix}
\\[24pt]

In all calculations we are using the light-cone representation
of a four-vector $V^\mu$, defined as
\bq 
V^A=(V^0+V_z,V^0-V_z,V_x+i V_y,V_x-i V_y)\;,\;\; A=1,\ldots,4\;. 
\eq
Polarization state-vectors are given by
\bqa
\epsilon_-^A&=&\left(
\frac{-p_T}{\sqrt{2}|\vec{p}|},
\frac{p_T}{\sqrt{2}|\vec{p}|},
\frac{(p_x+ip_y)(|\vec{p}|+p_z)}{\sqrt{2}|\vec{p}|p_T},
\frac{(p_x-ip_y)(-|\vec{p}|+p_z)}{\sqrt{2}|\vec{p}|p_T}\right)
\nn\\
\epsilon_+^A&=&\left(
\frac{p_T}{\sqrt{2}|\vec{p}|},
\frac{-p_T}{\sqrt{2}|\vec{p}|},
\frac{(p_x+ip_y)(|\vec{p}|-p_z)}{\sqrt{2}|\vec{p}|p_T},
\frac{(p_x-ip_y)(-|\vec{p}|-p_z)}{\sqrt{2}|\vec{p}|p_T}\right)
\nn\\
\epsilon_0^A&=&\left(
\frac{|\vec{p}|}{\sqrt{p^2}}+
             \frac{p_zp_0}{|\vec{p}|\sqrt{p^2}},
\frac{|\vec{p}|}{\sqrt{p^2}}-
             \frac{p_zp_0}{|\vec{p}|\sqrt{p^2}},
\frac{(p_x+ip_y)p_0}{|\vec{p}|\sqrt{p^2}},
\frac{(p_x-ip_y)p_0}{|\vec{p}|\sqrt{p^2}}\right)
\eqa
 
As for the Dirac matrices we are using the chiral 
representation~\footnote{For conventions see reference~\cite{itzy}.}.
The wave functions which describe massive spinors are given by:
\bqa
u_+(p)=\left(\begin{array}{c}
r/c\\
a(p_x+ip_y)/r\\
-mb/r\\
-m(p_x+ip_y)/r
\end{array} \right)
&&
\bar{u}_+(p)=\left(\begin{array}{c}
mb/r\\
m(p_x-ip_y)/r\\
-r/c\\
-a(p_x-ip_y)/r
\end{array} \right)
\nn\\   
u_-(p)=\left(\begin{array}{c}
m(p_x-ip_y)/r\\
-mb/r\\
-a(p_x-ip_y)/r\\
r/c
\end{array} \right)
&&
\bar{u}_-(p)=\left(\begin{array}{c}
a(p_x+ip_y)/r\\
-r/c\\
-m(p_x+ip_y)/r\\
mb/r
\end{array} \right)
\nn\\ 
v_+(p)=\left(\begin{array}{c}
-m(p_x-ip_y)/r\\
mb/r\\
-a(p_x-ip_y)/r\\
r/c
\end{array} \right)
&&
\bar{v}_+(p)=\left(\begin{array}{c}
a(p_x+ip_y)/r\\
-r/c\\
m(p_x+ip_y)/r\\
-mb/r
\end{array} \right)
\nn\\ 
v_-(p)=\left(\begin{array}{c}
r/c\\
a(p_x+ip_y)/r\\
mb/r\\
m(p_x+ip_y)/r
\end{array} \right)
&&
\bar{v}_-(p)=\left(\begin{array}{c}
-mb/r\\
-m(p_x-ip_y)/r\\
-r/c\\
-a(p_x-ip_y)/r
\end{array} \right)
\eqa
where:
\[
a=p_0+|\vec{p}|,\; \; b=p_z+|\vec{p}| ,\; c=2|\vec{p}|,\;r=\sqrt{abc}
\]
For a massless particle the spinors are
\bqa
u_R(p)=
\left(\begin{array}{c}
\sqrt{p_0+p_z}\\
(p_x+ip_y)/\sqrt{p_0+p_z}\\
0\\
0
\end{array} \right)
&&
\bar{u}_R(p)=
\left(\begin{array}{c}
0\\
0\\
-\sqrt{p_0+p_z}\\
-(p_x-ip_y)/\sqrt{p_0+p_z}
\end{array} \right)
\nn\\
u_L(p)=
\left(\begin{array}{c}
0\\
0\\
-(p_x-ip_y)/\sqrt{p_0+p_z}\\
\sqrt{p_0+p_z}
\end{array} \right)
&&
\bar{u}_L(p)
=\left(\begin{array}{c}
(p_x+ip_y)/\sqrt{p_0+p_z}\\
-\sqrt{p_0+p_z}\\
0\\
0
\end{array} \right)
\eqa
 
We now proceed to describe the vertex functions. Let us take as an
example the \eqn{vff}. By using the Dirac matrices in the chiral
representation and the light-cone expression of  four vectors, the
reduced form of this equation becomes very simple, namely the four
vector 
\[
V_\mu=\psb(P_1)\gamma_\mu\left(g_R\omega_R+g_L\omega_L\right)\psi(P_2)
\]
turns out to be 
\bq V^A =\left(\begin{array}{c} -g_R
\psi_1\psb_3- g_L \psi_4\psb_2 \\ -g_R \psi_2\psb_4- g_L
\psi_3\psb_1 \\ -g_R \psi_2\psb_3+ g_L \psi_4\psb_1 \\ -g_R
\psi_1\psb_4+ g_L \psi_3\psb_2
\end{array}
\right)
\eq
where $\psi_i(\psb_i), i=1,\ldots,4$ are the components of the spinor
$\psi(P_2)\left(\psb(P_1)\right)$ and
\bq
\omega_L=\frac{1}{2}\left(1-\gamma_5\right)\;,\;\;
\omega_R=\frac{1}{2}\left(1+\gamma_5\right)\;\;.
\eq
On the other hand, the spinor 
\[
u=(\slaa{P}{-7.5}+m)\slaa{b}{-5.5}(P_1)\omega_R\psi(P_2)
\]
used
for instance in \eqn{ffv}, can be reduced to 
\bq u
=\left(\begin{array}{c}
(-b_2 p_1+b_3p_4)\psi_1+(b_4p_1-b_1p_4)\psi_2 \\
(b_3 p_2-b_2p_3)\psi_1 +(-b_1p_2+b_4p_3)\psi_2 \\
m(b_2\psi_1-b_4\psi_2)\\ m(-b_3\psi_1+b_1\psi_2)
\end{array}
\right)
\eq
where $\psi_i,\;b_i,\;p_i,\;i=1,\ldots,4$ are the components of 
$P$, $b(P_1)$ and $\psi(P_2)$
respectively. In a similar way, for the
standard electroweak theory in both the unitary and the Feynman
gauge, twenty eight  different vertex functions have been implemented
in {\tt vertices/pan2.f}.

\vfill

\noindent{\Large \bf Acknowledgements} \\[24pt]
Helpful discussions with F.~A.~Berends, A.~P.~Chapovsky and R.~Pittau
are kindly acknowledged.

\end{document}